%% file: main.tex
\documentclass[a4padisinvitingper,fleqn]{latex_styles/cas-sc}
\input{latex_styles/preamble.tex}

\begin{document}
\let\WriteBookmarks\relax
\def\floatpagepagefraction{1}
\def\textpagefraction{.001}

\shorttitle{Sources of capital growth}

\shortauthors{Gordon Getty}

\title [mode = title]{Sources of capital growth}                      


%
\author[1]{Gordon Getty}[
                        type=editor,
                        role=Researcher,
                        orcid=0000-0002-0939-6932
                        ]

\ead{ggetty@LeakeyFoundation.org}

\author[2]{Nikita Tkachenko}[
                        type=editor,
                        role=Assistant,
                        orcid=0009-0003-8681-3335
                        ]

\cormark[1]

\ead{natkachenko@usfca.edu}




\affiliation[1]{organization={Fellow, UCB},
            addressline={University Avenue and, Oxford St}, 
            city={Berkeley},
            postcode={94720}, 
            state={California},
            country={United States}}
            
\affiliation[2]{organization={Graduate Student, University of San Francisco},
            addressline={2130 Fulton St}, 
            city={San Francisco},
            postcode={94117}, 
            state={California},
            country={United States}}





\cortext[cor1]{Corresponding author}



\begin{abstract}
Data from national accounts suggest that
net saving, at the collective scale, contributes less than 8\% to growth in capital value.
We explore ways in which this is
possible, and discuss implications for economic teaching and public
policy.     
\\ 
\\
\end{abstract}




\maketitle

\hypertarget{introduction}{%
\section{Introduction and overview}\label{introduction}}

Many economists over the centuries have reasoned that net saving\footnote{As reported in national accounts; saving and investment differ only by statistical discrepancy. We view them as different, for reasons shown in Appendix \hyperref[appendix-a]{A}, but will equate them in this paper by using the word "saving" in the Keynesian sense of invested saving.} should tend to give equal capital growth. Economists since the early
nineteenth century have added the proviso that net saving cannot
safely outpace innovation; more capital must mean capital redesigned for
greater productivity if economies are to escape risk of capital glut and
diminishing returns
(\citet{westEssayApplicationCapital1815, ricardoEssayProfitsVol1815, malthusEnquiryNatureProgress1815}).
Roy \citet{harrodEssayDynamicTheory1939}
described that limit for safe net saving, meaning the rate of
imagining and developing new ideas for more productive forms of capital,
as the ``warranted rate''. Harrod, and many other economists of his time
and since, have focused on growth of output rather than of capital, but
have modeled growth of output by first assuming the equivalence of net
saving and capital growth, within the warranted rate, and then
looking for effects of that capital growth on later output growth.

Some other economists, including John  \citet{raeNewPrinciplesPolitical1834} and John Stuart 
\citet{millPrinciplesPoliticalEconomy1848} (see Appendix \hyperref[appendix-c]{C}), argued that capital growth might also be explained
by a rise in productivity of capital and labor already extant. Ways
might be found for existing factors to produce more, that is, and so to
allow more consumption, or more capital growth, or any mix of the two,
without inputs of net saving. Robert
\citet{solowTECHNICALCHANGEAGGREGATE1957} allowed that possibility for
``disembodied'' growth, where plant and products already existing are
repurposed or redeployed in more productive ways.

We test between those two explanations of capital growth, by net
saving or by increase in productivity of capital and labor already
in existence, by comparing net saving to concurrent change in
market-value capital in 92 countries, 
and also comparing changes in those two variables.
All data are drawn from national accounts
of those countries as collated on the free website \href{https://wid.world/}{World Inequality Database}. Our findings 
imply that the efficiency of net saving in realizing capital growth has been less than 8\%. We will argue that it may be substantially less, and possibly zero.\footnote{We will find three measures, of which the highest is 7.71\%, of the causal relatedness between net saving and capital growth. These measures do not give the direction of causality. We will argue that exogenous 
    disruptions such as wars or epidemics or supply or demand shocks can bring market value and income down as a direct effect, and that households, in response, may be motivated to trim saving in order to maintain consumption. In these cases, the drop in capital growth \(g(K)\) would precede and explain the drop in net saving \(S_{net}\). The same direction of causality would appear in recoveries from these setbacks. As capital values rise again, that is, income would rise with them, and households would be motivated to restore old saving rates as insurance against the next downturn. (See Appendix \hyperref[appendix-e]{E}).
} Net saving, allowing for that margin of possible exception, has raised the physical quantity (volume) of capital, but not the
aggregate value, and so has reduced the value per unit.
This condition describes capital glut.

The explanation we suggest for capital growth without net saving is in production efficiencies of innovation. 
The innovator acquires materials and plant capacity and labor skills at market prices
determined by their uses in current technology, but applies them more
productively until competition catches up. That temporary
advantage to the innovator in market value of output per unit cost of input, which we call the innovator's reserve, creates new income from which to pay that cost, and leaves original income free to pay for consumption. This new income expands the economy, meanwhile, and justifies an equal expansion of the money supply to maintain price stability. Thus growth by a gain in productivity of existing resources costs no sacrifice in consumption, and no diversion of funds from other uses.

An innovator in need of outside funding may approach investors. Investors calculate present value, just as with projects relying on established rather than innovative technology, and invest up to that amount.\footnote{Present value is the sum of time-discounted cash flows expected. These flows tend to be negative over any period of investment, and then positive as products are finished and sold. An innovative project may need more investment than others, and a longer period of investment, and will not reward investment unless positive cash flows at the end, discounted to present value, are large enough to offset.}\textsuperscript{,}%
\footnote{An innovator not in need of outside funding would still require equal present value as a minimum condition for investment.}\textsuperscript{,}%
\footnote{Competition among investors should tend to keep investment close to present value.} Innovative projects may need more investment over a longer period than others, but will still justify funding if expected return at the end is seen as enough to bring present value equal at least to investment. Any excess of present value over investment quantifies the investor's reserve.\footnote{Appendix \hyperref[appendix-d]{D} will show that an innovative investment will produce cash flows forward at the rate of return \(r_I\) realized by the innovation, while those cash flows will be discounted backward to present value at the market time-discount rate \(r_m\) (see Appendix \hyperref[appendix-d]{D}).} 

We further suggest, without evidence,
that innovation should tend to bring positive
externalities to capital of third parties
not participating in the innovation. Old
structures and infrastructure, for example,
should tend to gain value as they house or 
support higher-tech workers and equipment.
We call this effect "tech externalities."
"Free growth" will mean the combined growth in capital
of innovators through the innovator's reserve,
and in capital generally through tech externalities.
Free growth theory will mean the prediction 
that the rate of productivity gain, which Harrod called the warranted rate, tends to bring an equal rate of capital growth through the innovator's reserve and tech externalities, without help from net saving, and so brings most or all of capital growth not explained by market noise. Thus the theory allows for some contribution by net saving, but expects any such contribution to be small. We will use the term "thrift residue" to mean any part of capital growth that might be explained by net saving, less than 8\% by our findings, with the understanding that the part actually explained by net saving may be zero.

Although net saving is expected to be largely or wholly ineffective, the theory does not question the value of gross saving to make up for losses to depreciation. If these losses are not made up first, value created by productivity gain will go first to make up this shortfall, and only the remainder will apply to capital growth.

The theory predicts only at the largest scales, where the data are clear, and only for the private sector.
We accept that households commonly grow or decline in net worth through net saving or dissaving (see Section \ref{disclaimers}), and that the same is possible for larger groups and even small nations. Free growth theory predicts 
only at the scale of all capital and all economies together. The public sector,
meanwhile, responds to political rather than market choices, and grows
or shrinks accordingly.

If free growth theory is right, tax policy and other policy to encourage
saving over consumption should be reviewed. These policies
include the higher tax on ordinary income than on capital gains, and the
double tax on corporate dividends.

Inferences for economic teaching include the obvious ones for growth
theory and for net saving in general. They include others as well.
One of the central doctrines of the marginalist revolution has held that
market realization converges to producer cost, when that cost includes
imputed interest on assets owned.
A gain in productive efficiency can bring market value above producer cost, and by most or all of capital growth if free growth theory holds.
Meanwhile the doctrine that net income equals
consumption plus net saving is put into question by evidence offered
here suggesting that net saving increases the physical quantity of
capital, but adds less than 8\% to aggregate value. 

\hypertarget{net-saving-and-capital-growth}{%
\section{Net saving, capital growth and capital acceleration}\label{net-saving-and-capital-growth}}

Thrift theory will mean the combined ideas that net saving \(S_{net}\) is realized in equal growth of capital \(K\), if \(S_{net}\) holds within the warranted rate, and that there is no other source of capital growth except market noise regressing to zero. "Thrift assumptions" will mean those two plus the assumption that current \(S_{net}\) holds within the warranted rate. The acronym uta will mean "under thrift assumptions." Then 
\begin{equation}
    \Delta K = S_{net}\ ,\quad \text{uta}. \label{eq-1}
\end{equation}
Define "thrift" \(s^*\) by \(s^* = \frac{S_{net}}{K}\), and divide Eq. \eqref{eq-1} by capital to predict capital growth rate \(g(K)\) as
\begin{equation}
g(K) = s^* \ , \quad \text{uta,}\quad \text{where } s^* = \frac{S_{net}}{K}.\quad \text{ Then} \label{eq-2}
\end{equation}
\vspace{-5ex}
\begin{equation}
\Delta g(K) = \Delta s^*\ , \quad \text{uta.\footnotemark} \label{eq-3}
\end{equation}
\footnotetext{The asterisk distinguishes \(s^*\) from the Keynesian saving rate \(s\), where the denominator is output rather than capital, and the numerator shows gross rather than net saving.}
\(\Delta g(K)\) and \(\Delta s^*\) respectively will be called "capital acceleration" and "thrift change," either of which may be positive or negative or zero. Division of Eq. \eqref{eq-3} by capital acceleration, with rearrangement, gives
\begin{equation}
\frac{\Delta s^*}{\Delta g(K)} = \frac{\Delta g(K)}{\Delta g(K)} = 1\ , \quad \text{uta}. \label{eq-4}
\end{equation}

Algorithms for correlation and covariance and regression are structured on the reasoning that causal relationships among variables tend to be revealed in concurrent changes. While Eqs. \eqref{eq-1} and \eqref{eq-2} enable measurement of the coexistence of net saving and capital growth, Eqs. \eqref{eq-3} and \eqref{eq-4} allow a better measure of the degree to which that coexistence is explained by causality.
Eq. \eqref{eq-4} was derived from Eq. \eqref{eq-3} so that success of predictions from thrift assumptions can be measured against a standard of unity. For notational convenience, then, we define the "thrift index" \(\theta\) by \(\theta = \frac{\Delta s^*}{\Delta g(K)}\), and restate Eq. \eqref{eq-4} as
\begin{equation}
\theta = 1\ , \quad \text{uta,}\quad \text{where} \ \theta = \frac{\Delta s^*}{\Delta g(K)}\ . \label{eq-5}
\end{equation}

Test results for the predictions \(\frac{s^*}{g(K)} = 1\ \text{and}\ \theta = 1\), under thrift assumptions, appear in Fig. \ref{fig-s_c_theta_plots} and in Tables \ref{tbl-reg_s}, \ref{tbl-5} and \ref{tbl-indicator_table}. Screens to which they refer are explained in Section \ref{sec-displays} below. Results were found as GDP-weighted averages over all countries and years, from data for net saving and market-value capital taken from national accounts. 
Figures \ref{fig-s_c_theta_plots} and \ref{fig-si_plots} show \(\frac{s^*}{g(K)}\) and \(\frac{\Delta s^*}{\Delta g(K)}\) as 0.481 and 0.064 respectively. Table \ref{tbl-4} shows regression of \(s^*\) on \(g(K)\) as 0.0771, and regression of \(\Delta s^*\) on \(\Delta g(K)\) as 0.0559.

\section{Interpretation of test results\label{interpretation-of-test-results}}

The finding \(\frac{s^*}{g(K)} = 0.481\) reveals average observed proportions between \(s^*\) and \(g(K)\), we repeat, and does not of itself show causal relatedness between those two variables. Causal relatedness is shown rather by the degree of constancy in proportions, and thus on regression of \(s^*\) on \(g(K)\) found at 0.0771. \(\theta\) or \(\frac{\Delta s^*}{\Delta g(K)}\), measured at \(0.064\), gives a separate measure of causal relatedness between \(s^*\) and \(g(K)\) as explained. Regression of \(\Delta s^*\) on \(\Delta g(K)\), found at 0.0559, gives a measure of causal relatedness between \(\Delta s^*\) and \(\Delta g(K)\). Might these small but positive findings allow the possibility that some capital growth is explained by net saving?

Appendix \hyperref[appendix-e]{E} will develop our argument in Section \ref{introduction} (footnote 2) that the causal relatedness revealed in \(\theta\) and its associated regressions might also be explained as joint effects of outside events affecting \(g(K)\) directly, and \(s^*\) indirectly in adjustment to those effects on \(g(K)\). We also note that investment in government-owned property such as infrastructure is excluded from predictions of free growth theory, but included in net investment reported in the national accounts from which we take our data. Much government-owned property is unsold, and impractical to appraise, and accordingly is valued in national accounts at depreciated cost. Thus net investment in such government assets shows in national accounts as equal to capital growth, and contributes to \(\theta\). Free growth theory takes no position on the actual role of net saving in growth of government-owned property, and consequently neglects this contribution to \(\theta\) as outside the scope of its prediction.

\section{Inferences for the creative destruction idea}\label{inferences-for-the-creative-destruction-idea}

Creative destruction (\cite{schumpeter1912}\footnote{Published by Schumpeter in 1912, and translated by Opie in 1934.}) is
obsolescence of existing capital through innovation of new and preferred
technology, as with horse-drawn carriages and automobiles. Granting the
obsolescence of old ways by new, it does not follow that innovation reduces
values of assets already present. We have argued that
capital growth is maximized when investment is enough to offset depreciation first. A natural source
of that investment is depreciation of assets the new technology will replace,
and a natural market for the new technology may be owners of the assets to be
replaced. Carriage owners were better able to afford cars when the carriages in
which they had invested  before lived out their economic lives as originally
expected, so that the depreciation component in the cash flows originally
expected can pay for cars in replacement just as it might have paid for new
carriages. 

This analysis suggests that market forces favor a kind of orderly obsolescence
through replacement, without premature displacement, as the new technology awaits its turn.

\FloatBarrier
\begin{table}[pos=H]
    \caption{Regression of \(s^*\) on \(g(K)\) and \(\Delta s^*\) on \(\Delta g(K)\), GDP-weighted. Screen = 0.01. \(H_0\) per thrift theory: \(\frac{s^*}{g(K)} \cong \frac{\Delta s^*}{\Delta g(K)} \cong 1\).\label{tbl-reg_s}}
    \hrule
    \centering
\begin{minipage}{0.45\textwidth}
\input{./tables/tbl-4.tex}
\end{minipage}
\hfill
    \begin{minipage}{0.45\textwidth}
\input{./tables/tbl-wid_si_table.tex}
\end{minipage}
\hrule
\end{table}
%
%
\begin{figure}[pos=H]
    \centering
        \includegraphics[width=0.9\textwidth]{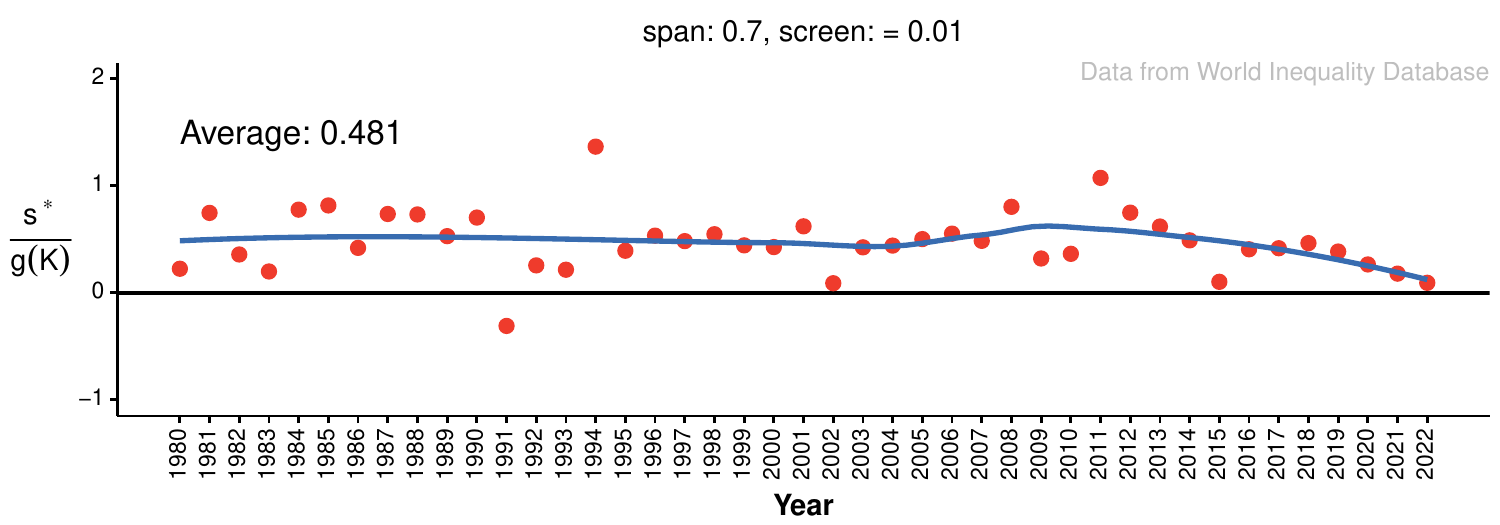}
    \captionsetup{justification=centering}
    \caption{\textcolor[HTML]{EF3B2C}{Average \(\frac{s^*}{g(K)}\)} with \textcolor[HTML]{386CB0}{LOESS smoothing} over all countries, GDP-weighted, 1980-2022} 
    \label{fig-s_c_theta_plots}
\end{figure}
%
%
\begin{figure}[pos=H]
    \centering
        \includegraphics[width=0.9\textwidth]{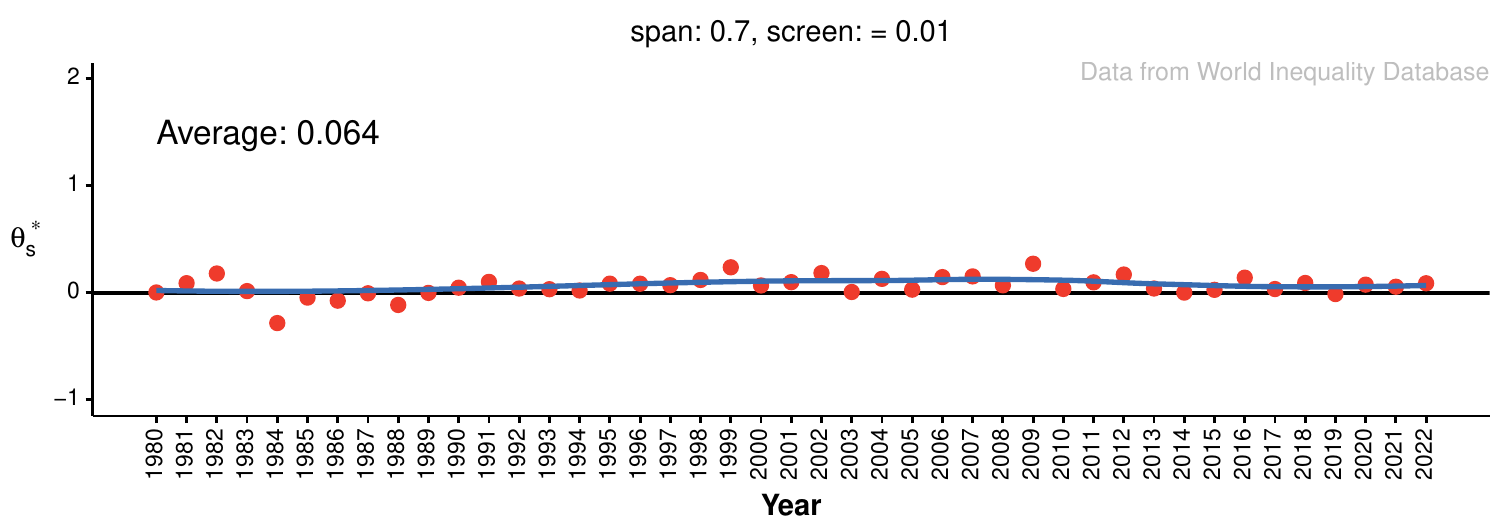}
    \captionsetup{justification=centering}
    \caption{\textcolor[HTML]{EF3B2C}{Average \(\theta_s^*\)} with \textcolor[HTML]{386CB0}{LOESS smoothing} over all countries, GDP-weighted, 1980-2022} 
    \label{fig-si_plots}
\end{figure}
\FloatBarrier
\input{./tables/tbl-5.tex}
\FloatBarrier

\FloatBarrier
\input{./tables/tbl-indicator_table.tex}
\FloatBarrier
\hypertarget{solows-puzzle}{%
\section{Solow's puzzle}\label{solows-puzzle}}

Some growth is capital widening, where structures and implements
increase in number but do not change in design. Capital widening,
however, is practical only so far before glut and diminishing returns
set in. Further growth must come from capital deepening,
meaning improvements in the design of capital.
\citet{solowContributionTheoryEconomic1956a} noted a kind of middle
ground between capital widening and capital deepening in the disembodied
growth mentioned earlier; ships carrying coals to Newcastle can raise
prospective cash flows, and hence present value, by reversing the
business plan. But Solow, who came to conclusions similar to ours from
different evidence, puzzled as to how capital growth without net
saving could be possible for capital deepening through ``embodied''
growth, where products of new design are made from plant of new
design.

Solow received the Nobel Prize in 1987. His acceptance speech mentions the embodiment problem:\footnote{\cite{solow1988}. The terms capital deepening, capital widening, embodied growth and disembodied growth are all Solow's.}

"... much technological progress, maybe most of it, could find its way into actual production only with the use of new and different capital equipment. Therefore the effectiveness of innovation in increasing output would be paced by the rate of gross investment. A policy to increase investment would therefore lead to ... a faster transfer of new technology into actual production ... By 1958 I was able to produce a model that allowed for the embodiment effect. If common sense was right, the embodiment model should have fit the facts better than the earlier one. But it did not... I do not know if that finding should be described as a paradox, but it was at least a puzzle."

Solow was right in that much innovation requires months or years of negative cash flow, in the form of gross investment, in building new plant before the new products roll out from the factory doors. From arguments in Section \ref{introduction}, however, this investment would not have been forthcoming unless positive cash flows at the end of the tunnel were predicted to be sufficient to create equal or greater present value from the start.(See Appendix \hyperref[appendix-d]{D}.) That initial creation of present value, plus replacement investment,\footnote{Also called depreciation investment.} are enough.

\hypertarget{mechanics-of-free-growth}{%
\section{Mechanics of free growth}\label{mechanics-of-free-growth}}

In terms of Solow's puzzle, we suggest that embodied growth is disembodied growth
on a finer scale. At each step toward realization of the new plant and
products, raw materials and products and labor skills and plant capacity
currently available on the market are adapted to new uses. The innovator
pays for these inputs at a market price determined by their value in
established productive uses, but applies them innovatively to realize
higher prospective cash flows, and hence higher present values, to the
innovator
(\citet{marshallPrinciplesEconomics1890, schumpeterTheoryEconomicDevelopment1934}).
That new use of materials and skills applied could give the solution to Solow's puzzle with respect to practical feasibility, as argued in Section \ref{introduction}, while the economic advantage to the investor, here called innovator's reserve,\footnote{i.e., reserve price which the innovator, who realizes more value from capital and labor inputs, might have been willing to pay for those resources.} could give the solution with respect to source of payment. Appendix \hyperref[appendix-d]{D} will clarify.

Our findings support those of \cite{picketyCapitalIsBack2014} and \cite{kurz2023market} as to the market power of innovators to explain capital growth beyond net saving. Those studies assumed that net saving is recovered in capital growth, so that this market power, meaning the innovator's reserve, would explain only the part of capital growth remaining. Again, we go farther by questioning the assumption that net saving contributes even a part of capital growth.

\hypertarget{tech-externalities}{%
\section{Tech Externalities}\label{tech-externalities}}

We argued in Section \ref{introduction} that innovation can add value eslewhere. Old factories and warehouses and residences gain in utility as they house higher-tech equipment and worker skills. Old highways do the same as they carry higher-tech freight and drivers and passengers to higher-tech destinations in higher-tech vehicles.\footnote{\cite{raeNewPrinciplesPolitical1834} makes essentially the same argument in terms of the technology of his time.}

Most externalities,\footnote{Externalities are defined as effects of events on nonparticipants.} positive and negative, are part of random market noise expected to offset to zero over time and scale. The examples above, which we call tech externalities, are exceptions in that they can impose a prevailingly positive vector on capital value resisting the downward pull of depreciation. Free growth, in the full sense, means effects of the innovator's reserve on capital of the innovator, plus effects of tech externalities on capital in general.

Note that Solow's capital deepening, or growth through more efficient use of resources, fits the definition of productivity gain affording the innovator's reserve. The distinction between capital deepening and widening is otherwise complex. Early investors in technological gain, say venture capitalists, may be rewarded with returns above the current norm. Such early backers form a kind of second-order innovators, and that advantage in return contributes to the innovator's reserve. Meanwhile other producers may learn and adopt the same technology in a form of externality called spillover, and thus earn above-average returns before the market catches up. Products made by those producers may tend to duplicate those made by the original innovator, but are innovative by comparison to products not yet adopting the innovation. Thus we do not seek to draw a clear line between capital deepening and capital widening, or between tech externalities and the innovator's reserve.

\section{Market value and book value}\label{market-value-and-book-value}

It is natural to reason that depreciation gives the only change in capital value, once investment has ended, apart from market noise converging to zero. Insofar as depreciation can be predicted, then, depreciated cost would show the norm (expectation) toward which current market value converges. It would follow that investors might anticipate the market by selling when prices are above depreciated cost, and buying when below.
This reasoning would justify the reliance of modern national accounts, and the equations of macroeconomics (macro) which they were designed to inform, since the inception of both in Keynes's time, on depreciated cost as the preferred measure of capital value. \footnote{As argued by James Tobin in \citeyear{tobin1968} and \citeyear{tobin1969}. "Tobin's \(q\)" is the ratio of market to replacement value (RV). RV means depreciated cost under replacement cost accounting, where past investments and subsequent depreciation are re-expressed in the cheaper dollars of today. Investors were advised to invest when Tobin's \(q\) is less than one, and divest or withhold investment conversely.}

We have argued that capital grows and declines with effects of the innovator's reserve and tech externalities, as well as investment and depreciation and market noise. Consider, however, a period when innovation is slight, so that depreciation and market noise give substantially the only changes in value of an investment once booked. It would follow that depreciated cost (of investment)
could find the norm described, even if free growth were absent and depreciation were predicted perfectly, only insofar as did original cost. An argument that optimal accounting can give a sounder measure of worth than market value would seem  to need a supporting argument that price and worth were closer together at the time assets were acquired or created, and depreciation bases set at their cost, than over the depreciation period following. Garbage in, garbage out.

We do not question the value of depreciation accounting, or the importance of the depreciation concept in economic understanding. But we suggest that current market value gives a more reliable measure of capital, when both are known, than an assumption-dependent projection from past market value subject to the same criticisms as current market value.

These reflections help explain our own reliance on market-value capital. They might also suggest rethinking of the logic by which macro and national accounts have preferred to reason from replacement cost when data for market-value capital are available.

\hypertarget{optimum-investment-policy}{%
\section{Optimum investment policy}\label{optimum-investment-policy}}

Data and arguments suggest that the optimum amount of
saving, at the global scale, is replacement saving to offset depreciation, and nothing
more. That would not mean 
depreciation found by the depreciated cost method, which equates net investment to capital growth at all scales, and which cannot measure the contributions of the innovator's reserve or tech externalities.
Up to a point, it should be
possible to analyze the composition of market-value capital, and to model
depreciation of the whole,
just as national balance sheet accounting does with replacement value capital today.
A better plan, as
\citet{solowContributionTheoryEconomic1956a} wrote in response to
Harrod's knife edge argument (Harrod, \citeyear{harrodEssayDynamicTheory1939}), is to
expect the market to maximize rate of return, and to sense the point
where glut begins and returns fall.\footnote{Harrod had argued that
saving must hit the warranted rate exactly or risk positive
feedback through the operation of the output/capital ratio
(accelerator).}
Markets do so imperfectly when tax and other public policy reward
saving over distributions and consumption. Findings in this paper
suggest reconsideration of such policies. These include the double tax on
dividends, and the greater tax rate on ordinary income than on capital
gains. Effects of removing the double tax, and removing the difference
between tax rates on ordinary income and on capital gains, could be
revenue-neutral and non-partisan if the corporate tax were raised to
match, if the tax rates on ordinary income and on capital gains met
somewhere between, and if thoughtful grandfathering eased the
transition.

\section{Optimum government policy}

Households dissave in consumption, while firms dissave through dividends and other distributions.
Public policy, in many nations, has inhibited dissaving in the belief that policies to promote net saving favor capital growth.
Rather they inhibit consumption, add less than 8\% to capital growth, and thus inhibit net output as a function of both together (see Appendix \hyperref[appendix-b]{B}).

The effect of repealing such policies would be to address capital glut not by investing less, but by disinvesting more.\footnote{Higher dividends and distributions by firms, that is, would enable higher consumption by households.} 
Demand for dividends should rise in response as shareholders are freed from the burden of the double tax, and firms would be incentivized to compete to meet that demand. As firms did so, and raised dividends, firms would find a lower share of income left to invest, and would tend to invest more selectively. Lower-return investment opportunities would tend to find fewer bidders, and average returns, in consequence, should tend to rise. Government might watch and wait, in such a scenario, and weigh other policy as outcomes appear.

\section{Inferences for microeconomics}

Microeconomics ("micro") studies the mutual effects of supply, demand and price. It teaches that producer cost, including interest as the cost of time, tends to equilibrate to market value realized. We have proposed an exception in the innovator's reserve; the innovator pays the going rates for skills and materials needed, and the going interest rate as the price of time, but applies them to make products of greater market value (See Appendix \hyperref[appendix-d]{D}).

\hypertarget{net-output}{
\section{Inferences for interpretation of national accounts}\label{net-output}
}
National accounts define the sum of consumption and investment as product. Common teaching identifies this sum, in its version net of depreciation, as net output in the sense of value added.
We have disputed this reasoning in our argument that net investment realizes less than 8\% of its cost in capital growth, inhibits consumption by its full cost, and thus tends to reduce the sum of both. We also critique it on another ground. Assume, for the sake of argument, that we are wrong, and that net investment realizes equal capital growth.
Appendix \hyperref[appendix-b]{B} will show an argument that to interpret value added as the sum of consumption and capital growth, and neither less nor more, would miss two components, one negative (human depreciation) and one positive (self-invested work), in value added to human capital. Both of these flows are unmeasurable by known means. Thus we have not sought to measure net output or its growth, and have treated capital growth alone.

Common teaching also equates the sum of investment and consumption to expenditure. We agree that spending on new goods and services can find no other dispositions, and that investment and consumption can find no other sources, if we recognize imputed payment for voluntary consumption services as in parenting.

By the argument in Appendix \hyperref[appendix-b]{B} noted above, consumption and market-value capital growth give the measurable components of true net output in the sense of value added. Then their sum, which might be thought of as market-value net product, would give a measurable proxy for net output, although not net output per se.

It appears from the above that gross product and what we call market-value net product are both informative. The first gives expenditure, although not value added, and the second gives at least a first approximation of value added.

\hypertarget{data-sources}{%
\section{Data sources}\label{data-sources}}

All our data are drawn from the free website \href{wid.world.com}{World
Inequality Database wid.world.com (WID)}. This source collates data from national accounts and tax data
of 105 countries in constant currency units, and adjusts them where needed to conform to current standards of the System of National Accounts
(SNA) published by the United Nations. We show results for the 92 of
those countries which report both factors, net
saving and market-value capital, needed for deriving
the thrift indexes. WID's source for these data is national accounts.

Net saving $S_{net}$ and market-value $K$ are taken from net national saving (msavin) and market-value Capital Wealth (mnweal) respectively. GDP, which we use only for weighting purposes in Figs. \ref{fig-s_c_theta_plots} and \ref{fig-si_plots}, is reproduced from GDP (mgdpro).

\hypertarget{accessing-our-results-and-methods}{%
\section{Accessing our results and
methods}\label{accessing-our-results-and-methods}}

Tables and other displays of our findings for each country, and showing
our methods of calculation, can be accessed at the
\href{https://web-appendix.shinyapps.io/Sources_of_Capital_Growth/}{web appendix (https://web-appendix.shinyapps.io/Sources\_of\_Capital\_Growth)}.

\hypertarget{sec-displays}{%
\section{Screening out small denominators}\label{sec-displays}}

Our displays include findings for \(\theta\) 
for each year in each country over the report period. These tend to show
upward and downward spikes in values of \(\theta\)
in some years. Those spikes tend to be associated with small absolute
values of denominators, in these cases \(\Delta g(K)\), in those
countries and years. Small denominators, even if measured perfectly, magnify errors in measurements
of numerators. Worse, when denominators are small, small mismeasurements in them might reverse the term in sign.

To maximize reliability of test results, we apply a range of screens to
omit years where absolute denominators fall below a given
threshold. Some displays in the web appendix show data for all years, regardless of denominator size.
Others screen out all years where
absolute denominators are less then 0.01, then 0.025, then continuing upward in
increments of 0.025 to a maximum screen of 0.15. \(\theta\)
is plotted for each country unscreened and at each of the
seven successive levels of screening.
The screen chosen in displays in this paper is 0.01.
The denominator whose absolute value is screened is capital acceleration
$\Delta g(K)$ or capital growth rate \(g(K)\) in all displays. 

Screening out years where absolute \(\Delta g(K)\) or $g(K)$ is small would cost
little in informative value even if measurements were exact. In those
years, there is little capital acceleration or capital growth, positive or negative, for
either thrift theory or free growth theory to explain. Market noise
alone might account for \(\Delta g(K)\) or $g(K)$ in such years. Screening reduces
the number of observations, but increases the reliability and
informative value of each.

\hypertarget{disclaimers}{%
\section{Disclaimers}\label{disclaimers}}

We accept that capital growth is impossible without capital replacement first, to make up losses to depreciation, and that the expected source of capital replacement is replacement investment. Thus we accept the necessity and efficacy of replacement investment. We dispute only the efficacy of net investment in realizing capital growth after replacement investment has made up for depreciation.

Evidence supports the proposal of \cite{modigliani1954utility}  that the young save to afford families and retirement, and that retirees dissave to pay for both as they arrive.\footnote{\cite{friedman1957permanent} follows a different argument to a similar conclusion.} Such age-based differences are expected to offset in the stationary state, where tradition and free growth theory agree that overall net saving is zero. We accept that individual net saving can realize equal individual capital growth, so long as collective saving does not exceed collective dissaving.

\hypertarget{discussion-and-conclusions}{%
\section{Discussion and conclusions}\label{discussion-and-conclusions}}

With disclaimers noted, and with allowance for uncertainties of the thrift residue, we have argued that the innovator's reserve and tech externalities pay for capital growth as a whole, and that net saving by some can realize equal capital growth only insofar as others dissave as much. Market appraisals raise present value at the warranted rate with no need for net saving at the overall scale. We adopted the Modigliani model, where young households save to afford families and retirement while older households dissave to pay for both as they arrive, as an example of this saving/dissaving balance.

These findings challenge the teachings that capital growth at all scales is effected
by net saving enabled by consumption restraint, and that producer
cost, including imputed interest as the cost of time,
converges to market realization. Evidence suggests that much or all capital growth at the collective scale is free, and
consequently that market value, in the presence of innovation,
exceeds producer cost by at least the majority of capital growth. Meanwhile the same evidence suggests a review of the teachings that consumption plus net saving gives net income, and that consumption plus net investment gives net output. Appendix \hyperref[appendix-b]{B} will also question the latter teachings.

Embodied growth is disembodied growth on a finer scale. It redeploys or
repurposes existing labor skills, raw materials, and plant capacity, as
well as existing finished goods, to achieve higher returns than
available from the customary uses which determine their prices. The
present value of yields from this advantage in return, or equivalently
the innovator's reserve, plus tech externalities, defines the non-random component in free growth.

Solow's puzzle runs deep. How can thoughts be things? How can assessments of 
cash flows over the future pay for their realization in the present, even when plant of new design must come first?
Science and philosophy push our minds into areas where they are not at home. Neither relativity nor quantum mechanics, which explore the scale of great speeds and minute distances, sink in easily. Economics, too, invites surprises as it looks farther from the scale of the household and community and firm to which we are accustomed. Free growth theory predicts only at the universal scale, and accepts that thrift explains a greater part of growth at progressively smaller scales. Our findings, with that allowance, are awkward for much of economics as generally taught. 

The data are compelling, even so, and our explanatory hypothesis in Section \ref{introduction} fits common observation. We can see that producing more from the same inputs can spare a need for cutting back elsewhere. And we know that markets invest according to forecast future earnings, even when present earnings are nil. Free growth theory adds only that these well-understood effects, described in \cite{picketyCapitalIsBack2014} and \cite{kurz2023market}, explain all or substantially all of capital growth.

\appendix
\renewcommand{\theequation}{A.\arabic{equation}}
\setcounter{equation}{0}

\hypertarget{appendix-a}{
\section*{Appendix A. \hspace{0.5em}Equivalence of saving and investment}\label{appendix-a}
}

\cite{keynesGeneralTheoryEmployment1936} argued correctly that saving, if not invested in currently produced producer goods and services, adds nothing to capital, and thus nothing to output. He inferred that uninvested saving has no value, as it accomplishes no growth, and described it as "intended saving."  True saving, by that reasoning, equalled saving actually invested.

We suggest that investors seek to maximize return, within risk tolerance, and will sometimes hold saving in cash or equivalents at zero return in recessions and depressions when positive returns cannot be found, and so when investing would tend to reduce capital and output rather than increase them.
Thus we do not share the Keynesian view that saving need be invested in order to maximize its value. Saving will tend to equal investment in normal times, when prospective returns bring animal spirits, and could include saving in cash when not.
It is only for these profitable times that we accept the equality of saving and investment, and only as a simplifying assumption.\footnote{The weakness in the Keynes argument, if we are right, was his inference that net saving finds no utility, and hence no economic value, except as invested to realize capital growth. Its utility in a more general sense, we suggest, is to maximize capital growth according to investment opportunities, and to maximize it by withholding investment when all investments promise loss.}

\renewcommand{\theequation}{B.\arabic{equation}}
\setcounter{equation}{0}

\hypertarget{appendix-b}{%
\section*{Appendix B.\hspace{0.5em}Net output with human
capital}\label{appendix-b}}

Human capital is impractical to measure, as it leaves little market
record other than for its rental income in pay and investment cost in nurture and schooling. Thus national accounts
leave it implicit, and allow us to infer what we can from data for pay and nurture and schooling. Those
accounts are founded on the principle, sound when terms are appropriately redefined, that net output,
or value added, is expressed in the sum of capital growth and net
outflow from the value-added chain. In national accounts, then, where
physical capital is the whole of capital while net outflow of the chain
is the whole of consumption, the reasoning is
\begin{equation}
Y = \Delta K + C \ , \quad \text{neglecting human capital.}\label{eq-a1}
\end{equation}
It is possible in principle to model a value-added chain which includes human
capital, and to compare findings with those shown in Eq.~\eqref{eq-a1}.
Let human capital $H$, in that new model, stand as the last link in the
value-added chain. Adapting the classic illustration of the value added
principle, that is, where farms produce wheat, mills convert the wheat to
flour, and bakeries convert the flour into bread, say that humans convert some of
the bread, called invested consumption, into human capital. The net
outflow from this extended value-added chain is not all of consumption,
but only pure consumption, meaning the part remaining after the part invested in human capital (invested consumption) is
subtracted.\footnote{The concepts of invested consumption, pure consumption, self-invested work and human depreciation were introduced in \cite{schultzInvestmentHumanCapital1961}.} By this reasoning, the principle
that net output is expressed in capital growth plus net outflow gives
\begin{equation} 
Y = \Delta K + \Delta H + C_{p}\ , \quad \text{allowing human capital,}\label{eq-a2} 
\end{equation}
where \(C_{p}\) gives pure consumption.

Yoram \citet{ben-porathProductionHumanCapital1967} reasoned
that growth in human capital equals invested consumption plus
self-invested work less human depreciation\footnote{Equation 4 in
  Ben-Porath's paper, summarizing his first three equations. His terms and
  notation differ from ours. He expressed his idea in Cobb-Douglas functions, whose values vary with scalars and exponents chosen, as opposed to our simpler algebra. We believe, however, that meanings agree.}.
Let \(C_{s},\ W_{s}\) and \(D(H)\) show these flows respectively.
Thus the combined arguments of Schultz and Ben-Porath arrive at
\begin{equation}
C = C_{s} + C_{p} \quad \text{and}
\quad \Delta H = C_{s} + W_{s} - D(H)\ , \quad \text{allowing human capital.}\label{eq-a3}
\end{equation}
Substitution of these equations into Eq.~\eqref{eq-a2} finds
\[Y = \Delta K + C_{s} + W_{s} - D(H) + C_{p} \quad \text{and consequently}\]
\vspace{-5ex}
\begin{equation}
    Y = \Delta K + C + W_{s} - D(H)\ , \quad \text{allowing human capital,}\label{eq-a4}
\end{equation}
if Schultz and Ben-Porath and the reasoning here are right.

\renewcommand{\theequation}{C.\arabic{equation}}
\setcounter{equation}{0}

\hypertarget{appendix-c}{%
\section*{Appendix C.\hspace{0.5em}Mill's statement of the free growth idea}\label{appendix-c}}

\cite{millPrinciplesPoliticalEconomy1848}, book 1, chapter 5, section 4, includes:
\begin{quote}
If it were said, for instance, that the only way to accelerate the increase of capital is by increase of saving, the idea would probably be suggested of greater abstinence, and increased privation. But it is obvious that whatever increases the productive power of labour creates an additional fund to make savings from, and enables capital to be enlarged not only without additional privation, but concurrently with an increase of personal consumption. 
\end{quote}

This passage may be the first clear statement of what we call free growth theory. Mill's use of the words "accelerate" and "concurrently" suggest that his path of reasoning was something like ours.

\renewcommand{\theequation}{D.\arabic{equation}}
\setcounter{equation}{0}

\hypertarget{appendix-d}{%
\section*{Appendix D.\hspace{0.5em}Present value and cost}\label{appendix-d}}

We use the term cash flow to mean value transferred from an asset to its owner, in cash or in kind, less value transferred from the owner to the asset. Thus cash flow will generally be negative over any period of investment by owners in buying or building the asset, and then positive when investment ends and net yields begin. Present value of the asset, at any moment, is the sum of future cash flows expected from that moment on, each discounted at the market time discount rate (see Appendix \hyperref[appendix-f]{F}). 
Thus the present value of a differential amount of future cash flow expected at time \(t\), shown as \(d F(t)\), if discounted at the market rate \(r_{m}\), finds a present value \(d PV\) given by
\begin{equation}
    d F(t) e^{-r_{m} t} = d PV \ .\label{eq-d1}
\end{equation}
Present value of all cash flows together, then, is
\begin{equation}
    PV = \int^{\infty} F(t) e^{-r_{m} t} d t\ ,\label{eq-d2}
\end{equation}
where the implicit lower limit of integration is the present time zero.

The actual investment needed to realize this present value will vary inversely with the productivity of the investment. Productivity of a given investment is measured in the rate of total return \(r_I\) it generates, composed of growth rate \(g(K)\) and cash flow rate. Consider a differential amount of investment \(d I\) invested now, and yielded out wholly, including both principal and accumulated growth, in a differential amount of cash flow \(d F\) at a single later time \(t\).\footnote{\(d I\) and \(d F\) need not represent the whole of investment or cash flow at times \(t_1\) and \(t_2\). It is always possible, in principle, to divide a given investment, including appreciation, into infinitesimal (differential) parts each yielded out at a single later moment.} As no part of principal or accumulated growth from this particular investment was yielded during the interim \(t\), by assumption, growth rate \(g(K)\), over that interim, equalled the whole of rate of return \(r_I\). Hence
\begin{equation}
    d F(t) = d I e^{r_{I} t}\ , \quad \text{or equivalently} \quad d I = d F(t) e^{-r_{I} t} \ .\label{eq-d3}
\end{equation}
Investment \(I\) is the sum of these differential components \(d I\). That is,
\begin{equation}
I = \int^{\infty} d I = \int^{\infty} d F(t) e^{-r_{I}t} d t \ .\label{eq-d4}
\end{equation}
Consider the case where \(r_I\) and \(r_m\) are equal. By Eqs. \eqref{eq-d1} through \eqref{eq-d4},
\begin{equation}
    dPV = d Fe^{-r_{m} t} = d F e^{-r_{I} t} = d I \quad \text{and} \quad PV = I\ , \quad  \text{if}\quad r_I = r_m \ .\label{eq-d5}
\end{equation}

Now relax that assumption, and allow \(r_m\) and \(r_I\) to differ. The innovator's reserve is defined as present value less cost invested. Thus the component in each differential cash flow due to the innovator's reserve is
\begin{equation*}
    d PV - dI = d F(t) e^{-r_m t} - dF(t)e^{-r_I t} = d F(t) (e^{-r_m t} - e^{-r_I t}) \ .
\end{equation*}
The innovator's reserve as a stock, which we show as \(RSV\), is accordingly found at
\begin{equation}
    RSV = PV - I = \int^{\infty} F(t) e^{-r_m t} d t - \int^{\infty} F(t) e^{-r_I t} d t = \int^{\infty} F(t) (e^{-r_m t} - e^{-r_I t}) d t \ .\label{eq-d6}
\end{equation}

These equations confirm that present value equals investment needed when \(r_I\) equals \(r_m\), and exceeds investment needed when \(r_I\) exceeds \(r_m\). 

\renewcommand{\theequation}{E.\arabic{equation}}
\setcounter{equation}{0}

\hypertarget{appendix-e}{%
\section*{Appendix E.\hspace{0.5em}Analysis of the thrift residue}\label{appendix-e}}

The data shown in our displays find the thrift residue by three measures ranging from 7.71\% to 5.59\%. This residue gives the extent to which \(s^*\) and \(g(K)\) are causally related, but does not give the direction of causality. It shows the extent to which change in \(s^*\) brought about a change in \(g(K)\), or conversely, or both were caused primarily by a common third factor. In the last case, the third factor might bring about a direct effect on \(s^*\) or \(g(K)\) alone, at first, which might then cause a reaction by the other. Thrift theory predicts that a change in \(s^*\) will bring an equal change in \(g(K)\) before allowance for market noise, whether the change in \(s^*\) was itself the primary cause, or whether it was a proximate cause driven by an exogenous shock, say war or epidemic or other disruption in supply or demand, as the primary one.

Internal misjudgement can contribute to these setbacks by weakening defences, or can bring them about without need for significant outside causes. Some economists attribute at least some recessions to underspending on currently produced consumer and producer goods, made worse by stock market skittishness. This interpretation would fit thrift theory in picturing a shortfall in net investment as causing the drop in market value of capital.

Another view pictures recessions as caused or facilitated by accumulated overinvestment through overestimation of growth prospects, and pictures recession at the end as a reality check. That scenario would reverse the direction of causality expected in thrift theory by putting the drop in \(g(K)\) and income first, and by seeing the drop in \(s^*\), the ratio of net saving to capital, as driven by a need to maintain consumption as income falls. 
The same chain of causality would repeat in the opposite direction as recovery from these crises  brings capital value and income up as a direct effect, and so enables households to save against the next crisis.

This issue was debated famously by Keynes and the Austrian school during the great depression, by Malthus, Say and Ricardo two centuries ago, and Smith and Mandeville before. They did not have the data reported in our displays. We suggest that the thrift theory explanation might have been urged less confidently in the light of information that net investment has accounted for less than 8\% of capital growth.

Other points can be added. Why should investors trade less efficiently in market equities than in the goods bought and sold by the underlying firms? This is the question we posed in our discussion of Tobin's \(q\) in Section \ref{market-value-and-book-value}.

\renewcommand{\theequation}{F.\arabic{equation}}
\setcounter{equation}{0}

\hypertarget{appendix-f}{%
\section*{Appendix F.\hspace{0.5em}Clarifying cash flow and \(r_m\)}\label{appendix-f}}

The term "cash flow" is not much used in economics outside the specialty of finance, where it is taken as the flow discounted to find present value, or added to capital appreciation to find total return. We extend this concept to non-financial as well as financial assets by interpreting the
services or utilities for which we buy a product as cash flows, or net withdrawals of value from assets to owners.\footnote{We buy a non-financial asset, say a book or chair or car or house, for the services (taste satisfactions) we expect from it, each discounted at the risk-adjusted time discount rate we each apply. As each expected satisfaction eventuates, we deduct its value from the pool of remaining expected satisfactions, and so from remaining value to us of the asset. Although we do these calculations intuitively rather than formally, they fit the essential criteria for of finding present value as the sum of time-discounted projected cash flows.} These cash flows are discounted to present value at the risk-adjusted time preference rate which the market applies to each product. Free growth theory predicts only at the universal scale, and is interested primarily in cash flows and time-discount rates at the scale of all products and all capital together. Thus we define \(r_m\), except where stated otherwise, as the market time-discount rate, or equivalently the cap-weighted average of risk-adjusted rates, for the economy overall.

Appendix \hyperref[appendix-d]{D} argued that \(r_m\) also gives the market-wide rate of return in the absence of innovation, but that productivity gain through innovation can bring rate of return above the time discount rate. Thus \(r_m\) can also be understood as the current standard rate of return before allowance for innovation. Note also that diminishing returns brought by capital glut should tend to drive \(r_m\) down, as the measure of overall return before effects of current innovation, and so make further investment easier to justify by enhancing present value under Eq. \eqref{eq-d2}. In this sense, capital glut may tend to be self-perpetuating.

\printcredits

\bibliographystyle{cas-model2-names}

\bibliography{cas-refs}

\end{document}

%% file: latex_styles/preamble.tex
\usepackage[authoryear,longnamesfirst]{natbib}

\usepackage{subfigure}
\usepackage{letltxmacro}
\hypersetup{
  colorlinks=false,
  hidelinks=false,
  pdfborder={0 0 0},
  pdfstartview={FitH}
}
\LetLtxMacro\oldhref\href
\renewcommand{\href}[2]{\oldhref{#1}{\uline{#2}}}
\usepackage{amsmath}
\usepackage{booktabs}
\usepackage{longtable}
\usepackage{array}
\usepackage{multirow}
\usepackage{wrapfig}
\usepackage{colortbl}
\usepackage{pdflscape}
\usepackage{tabu}
\usepackage[normalem]{ulem}
\usepackage{makecell}
\usepackage{xcolor}
\usepackage{caption}
\usepackage{tabularx}
\usepackage{adjustbox}
\usepackage{natbib}
\usepackage{mfirstuc}
\usepackage{float}
\usepackage{placeins}
\usepackage{footnote}
\bibpunct{(}{)}{,}{a}{}{;}

\DeclareCaptionFormat{bold}{\textbf{#1#2}#3}
\setcitestyle{aysep={}}
\setcitestyle{authoryear,open={(},close={)}}

\def\tsc#1{\csdef{#1}{\textsc{\lowercase{#1}}\xspace}}
\tsc{WGM}
\tsc{QE}
\tsc{EP}
\tsc{PMS}
\tsc{BEC}
\tsc{DE}


%% file: tables/tbl-4.tex
\centering
\begin{tabularx}{\columnwidth}{lcc}
   Regression of $s^*$ on \(g(K)\)   & 0.0771$^{***}$\\   
                                     & (0.0144)\\   
   Observations                      & 1,826\\  
   R$^2$                             & 0.84511\\  
   Within R$^2$                      & 0.10669\\  
    \\
   Year fixed effects                & $\checkmark$\\   
   Country fixed effects             & $\checkmark$\\   
\end{tabularx}
   \label{tbl-4}

%% file: tables/tbl-wid_si_table.tex
\centering
\begin{tabularx}{\columnwidth}{lcc}
   Regression of $\Delta s^*$ on $\Delta g(K)$   & 0.0559$^{***}$\\   
                                                 & (0.0057)\\   
   Observations                                  & 1,574\\  
   R$^2$                                         & 0.37071\\  
   Within R$^2$                                  & 0.27233\\  
    \\
   Year fixed effects                            & $\checkmark$\\   
   Country fixed effects                         & $\checkmark$\\   
\end{tabularx}
   \label{tbl-wid_si_table}

%% file: tables/tbl-5.tex
\begin{table}[pos=h]
\caption{Average \(\frac{s^*}{g(K)}\) in 92 countries over periods shown. Screen = 0.01. Number of years clearing screen shown in ()}\label{tbl-5}%
\hrule
\makebox[\textwidth][c]{%
{\centering

\begin{tabular}{llrr} Armenia & 1996 - 2020 (23) & -0.63\\
Aruba & 1996 - 2001 (6) & 1.25\\
Australia & 1980 - 2018 (33) & 0.24\\
Austria & 1996 - 2021 (23) & 0.83\\
Azerbaijan & 1996 - 2020 (23) & 1.57\\
\addlinespace
Bahrain & 2009 - 2013 (5) & -2.43\\
Belgium & 1996 - 2021 (21) & 0.61\\
Bolivia & 1998 - 2015 (17) & 0.40\\
Botswana & 1996 - 1999 (4) & 3.54\\
Brazil & 1996 - 2019 (21) & 0.25\\
\addlinespace
British Virgin Islands & 1996 - 1999 (4) & 2.45\\
Bulgaria & 1996 - 2016 (16) & 0.14\\
Burkina Faso & 2000 - 2019 (20) & 0.23\\
Cameroon & 1996 - 2019 (24) & 1.10\\
Canada & 1980 - 2022 (38) & 0.37\\
\addlinespace
Cape Verde & 2008 - 2017 (9) & 0.53\\
Chile & 1997 - 2021 (19) & 0.84\\
China & 1992 - 2020 (29) & 0.81\\
Colombia & 1996 - 2022 (27) & 0.60\\
Costa Rica & 2013 - 2020 (6) & 0.21\\
\addlinespace
Croatia & 1996 - 2021 (24) & 0.43\\
Curaçao & 2001 - 2016 (14) & 1.26\\
Cyprus & 1996 - 2021 (25) & 0.07\\
Czechia & 1994 - 2020 (20) & 0.17\\
Côte d’Ivoire & 1996 - 2000 (5) & 0.29\\
\addlinespace
Denmark & 1996 - 2022 (26) & 0.22\\
Dominican Republic & 1996 - 2016 (10) & 1.52\\
Ecuador & 2008 - 2020 (13) & 0.53\\
Egypt & 1997 - 2015 (19) & 0.92\\
El Salvador & 2015 - 2019 (5) & -0.62\\
\addlinespace
Estonia & 1996 - 2021 (24) & 0.45\\
Finland & 1996 - 2021 (23) & 0.17\\
France & 1980 - 2021 (35) & 0.26\\
Germany & 1980 - 2022 (38) & 0.70\\
Greece & 1996 - 2021 (26) & 0.17\\
\addlinespace
Guatemala & 2002 - 2021 (20) & -0.72\\
Guinea & 2004 - 2010 (5) & 0.72\\
Honduras & 2001 - 2015 (14) & -0.57\\
Hong Kong & 1997 - 2011 (15) & 0.90\\
Hungary & 1996 - 2021 (21) & 0.14\\
\addlinespace
Iceland & 2001 - 2014 (14) & 0.22\\
India & 1999 - 2019 (21) & 0.57\\
Indonesia & 2017 - 2019 (3) & 0.94\\
Iran & 1997 - 2017 (20) & 4.96\\
Ireland & 1996 - 2021 (22) & 0.26\\
\addlinespace
Israel & 1996 - 2019 (24) & 0.46\\ \end{tabular}
}

{\centering
\begin{tabular}{llrr} Italy & 1980 - 2022 (36) & 0.13\\
Japan & 1980 - 2021 (29) & 0.19\\
Kazakhstan & 1997 - 2022 (25) & 0.60\\
Kuwait & 2003 - 2017 (14) & 1.11\\
Kyrgyzstan & 1996 - 2021 (24) & 0.46\\
\addlinespace
Latvia & 1996 - 2021 (24) & -0.23\\
Lithuania & 1996 - 2021 (24) & 0.13\\
Luxembourg & 1996 - 2021 (22) & 0.03\\
Malaysia & 2007 - 2015 (9) & 2.52\\
Malta & 1996 - 2021 (23) & 0.16\\
\addlinespace
Mauritius & 2014 - 2018 (5) & 0.04\\
Mexico & 1996 - 2021 (22) & 0.15\\
Moldova & 1996 - 2019 (22) & -0.91\\
Mongolia & 2006 - 2020 (15) & 0.10\\
Morocco & 1999 - 2021 (23) & 1.13\\
\addlinespace
Netherlands & 1996 - 2021 (23) & 0.51\\
New Zealand & 1996 - 2019 (21) & 0.78\\
Nicaragua & 2007 - 2018 (12) & 0.08\\
Niger & 1996 - 2019 (24) & 0.90\\
Norway & 1981 - 2021 (38) & 0.95\\
\addlinespace
Peru & 2008 - 2021 (14) & 0.76\\
Philippines & 1996 - 2022 (27) & 1.76\\
Poland & 1996 - 2021 (25) & 0.71\\
Portugal & 1996 - 2022 (24) & 0.00\\
Qatar & 2002 - 2017 (14) & 1.89\\
\addlinespace
Romania & 1996 - 2020 (22) & 0.03\\
Russia & 1996 - 2017 (13) & -0.19\\
Saudi Arabia & 2003 - 2009 (7) & 2.65\\
Senegal & 2015 - 2021 (7) & 0.24\\
Serbia & 1998 - 2021 (23) & 0.61\\
\addlinespace
Slovakia & 1996 - 2022 (24) & 0.15\\
Slovenia & 1996 - 2021 (20) & 0.28\\
South Africa & 1996 - 2022 (24) & 0.28\\
South Korea & 1996 - 2020 (24) & 0.63\\
Spain & 1995 - 2021 (25) & 0.32\\
\addlinespace
Sweden & 1980 - 2021 (37) & 0.41\\
Switzerland & 1993 - 2021 (25) & 0.69\\
Tunisia & 1996 - 2011 (16) & 0.29\\
Turkey & 2010 - 2017 (8) & 1.32\\
USA & 1980 - 2021 (38) & 0.31\\
\addlinespace
Ukraine & 1996 - 2019 (21) & 0.17\\
United Kingdom & 1981 - 2021 (36) & 0.16\\
Uruguay & 2016 - 2017 (2) & 1.16\\
Uzbekistan & 2011 - 2021 (10) & 0.21\\
Vanuatu & 2002 - 2007 (6) & 0.29\\
\addlinespace
Venezuela & 1998 - 2019 (20) & 0.55\\ \end{tabular}

}
}
\hrule
\begin{flushleft}
\footnotesize \emph{Note:} \(s^*\) is defined as \(\frac{S_{net}}{K}\). Thrift theory predicts \(\frac{s^*}{g(K)} \cong 1\). Free growth theory makes no prediction for these data.
\end{flushleft}
\end{table}

%% file: tables/tbl-indicator_table.tex
\begin{table}[H]
\caption{Average \(\theta\) in 92 countries over periods shown. Screen = 0.01. Number of years clearing screen shown in ()}%
\hrule
\makebox[\textwidth][c]{%
{\centering

\begin{tabular}{llrr}
Armenia & 1997 - 2020 (20) & 0.12\\
Aruba & 1997 - 2001 (5) & 1.22\\
Australia & 1974 - 2019 (35) & 0.03\\
Austria & 1997 - 2021 (16) & 0.10\\
Azerbaijan & 1997 - 2020 (23) & 0.75\\
\addlinespace
Bahrain & 2010 - 2013 (4) & 0.41\\
Belgium & 1997 - 2021 (19) & 0.05\\
Bolivia & 1999 - 2014 (12) & 0.04\\
Botswana & 1997 - 2000 (4) & 0.24\\
Brazil & 1998 - 2020 (22) & 0.10\\
\addlinespace
British Virgin Islands & 1997 - 1998 (2) & 2.24\\
Bulgaria & 1997 - 2017 (15) & -0.03\\
Burkina Faso & 2001 - 2019 (16) & 0.04\\
Cameroon & 1997 - 2019 (19) & 0.18\\
Canada & 1974 - 2022 (41) & 0.09\\
\addlinespace
Cape Verde & 2009 - 2016 (8) & 0.27\\
Chile & 1998 - 2021 (22) & 0.06\\
China & 1981 - 2020 (33) & 0.07\\
Colombia & 1997 - 2022 (18) & -0.03\\
Costa Rica & 2014 - 2020 (6) & 0.04\\
\addlinespace
Croatia & 1997 - 2021 (22) & 0.19\\
Curaçao & 2002 - 2016 (14) & 0.41\\
Cyprus & 1997 - 2021 (23) & -0.06\\
Czechia & 1995 - 2021 (19) & -0.03\\
Côte d’Ivoire & 1997 - 2000 (4) & -0.03\\
\addlinespace
Denmark & 1997 - 2022 (22) & 0.05\\
Dominican Republic & 2007 - 2015 (9) & 0.15\\
Ecuador & 2009 - 2020 (12) & 0.17\\
Egypt & 1998 - 2015 (16) & 0.13\\
El Salvador & 2016 - 2019 (4) & 0.98\\
\addlinespace
Estonia & 1997 - 2021 (19) & 0.06\\
Finland & 1997 - 2022 (21) & -0.02\\
France & 1972 - 2021 (33) & 0.09\\
Germany & 1972 - 2022 (31) & 0.05\\
Greece & 1997 - 2021 (24) & 0.24\\
\addlinespace
Guatemala & 2003 - 2021 (11) & 0.15\\
Guinea & 2005 - 2010 (6) & 0.08\\
Honduras & 2003 - 2015 (10) & 0.40\\
Hong Kong & 1997 - 2011 (13) & 0.04\\
Hungary & 1997 - 2021 (21) & 0.07\\
\addlinespace
Iceland & 2002 - 2014 (12) & 0.10\\
India & 1997 - 2019 (16) & 0.14\\
Indonesia & 2019 - 2019 (1) & -0.41\\
Iran & 1998 - 2018 (19) & 1.09\\
Ireland & 1997 - 2021 (23) & 0.02\\
\addlinespace
Israel & 1997 - 2017 (18) & 0.07\\
\end{tabular}
}

{\centering
\begin{tabular}{llrr}
Italy & 1972 - 2021 (29) & 0.05\\
Japan & 1972 - 2021 (38) & 0.05\\
Kazakhstan & 1997 - 2022 (23) & 0.08\\
Kuwait & 2005 - 2017 (11) & 0.04\\
Kyrgyzstan & 1998 - 2021 (22) & 0.23\\
\addlinespace
Latvia & 1998 - 2021 (20) & 0.01\\
Lithuania & 1997 - 2021 (18) & 0.10\\
Luxembourg & 1997 - 2021 (25) & 0.00\\
Malaysia & 2008 - 2015 (7) & 0.01\\
Malta & 1997 - 2021 (23) & 0.08\\
\addlinespace
Mauritius & 2015 - 2018 (4) & 0.02\\
Mexico & 1997 - 2021 (22) & 0.06\\
Moldova & 1997 - 2019 (21) & 0.11\\
Mongolia & 2008 - 2021 (11) & 0.10\\
Morocco & 2000 - 2021 (17) & 0.19\\
\addlinespace
Netherlands & 1997 - 2021 (20) & 0.03\\
New Zealand & 1997 - 2019 (18) & 0.14\\
Nicaragua & 2008 - 2019 (12) & 0.06\\
Niger & 1997 - 2019 (19) & 0.22\\
Norway & 1982 - 2021 (34) & 0.12\\
\addlinespace
Peru & 2009 - 2021 (11) & 0.04\\
Philippines & 1997 - 2022 (20) & 0.37\\
Poland & 1997 - 2021 (15) & 0.20\\
Portugal & 1997 - 2022 (17) & 0.02\\
Qatar & 2004 - 2018 (12) & 0.29\\
\addlinespace
Romania & 1997 - 2020 (14) & 0.02\\
Russia & 1997 - 2018 (14) & 0.09\\
Saudi Arabia & 2005 - 2009 (3) & 2.23\\
Senegal & 2018 - 2021 (3) & 0.19\\
Serbia & 1999 - 2021 (22) & 0.22\\
\addlinespace
Slovakia & 1997 - 2021 (18) & 0.09\\
Slovenia & 1997 - 2021 (20) & 0.10\\
South Africa & 1997 - 2022 (21) & 0.10\\
South Korea & 1997 - 2020 (15) & 0.04\\
Spain & 1972 - 2021 (42) & 0.06\\
\addlinespace
Sweden & 1974 - 2022 (41) & 0.02\\
Switzerland & 1993 - 2020 (24) & 0.08\\
Tunisia & 1997 - 2011 (12) & 0.32\\
Turkey & 2011 - 2015 (5) & -0.16\\
USA & 1972 - 2021 (44) & 0.02\\
\addlinespace
Ukraine & 1997 - 2020 (23) & 0.05\\
United Kingdom & 1972 - 2021 (42) & 0.02\\
Uruguay & 2017 - 2017 (1) & 0.56\\
Uzbekistan & 2012 - 2021 (9) & 0.05\\
Vanuatu & 2003 - 2007 (3) & 0.15\\
\addlinespace
Venezuela & 1999 - 2019 (20) & 0.31\\
\end{tabular}

}

}
\hrule
\label{tbl-indicator_table}
\begin{flushleft}
\footnotesize \emph{Note:} \(\theta\) is defined as \(\frac{\Delta s^*}{\Delta g(K)}\). Thrift theory predicts \(\theta \cong 1\). See Section \ref{interpretation-of-test-results} for predictions of free growth theory. 

\end{flushleft}
\end{table}